\documentclass{article-jds}
\usepackage{graphics}
\usepackage{hhline}
\input{psfig.sty}
\input epsf

\nofirstpagebreak 

\begin{document}

\submitted{XXXV\`{e}mes Journ\'{e}es de Statistique, Sessions sp\'{e}ciales EFD}{1}

\title[Vers l'auto-administration des entrep\^{o}ts...]%
      {Vers l'auto-administration \newline des entrep\^{o}ts de donn\'{e}es}

\subtitle{}
\author{Kamel Aouiche \andauthor J\'{e}r\^{o}me Darmont}

\address{
Laboratoire ERIC, \'{E}quipe BDD\\
Universit\'{e} Lumi\`{e}re -- Lyon 2\\
5 avenue Pierre Mend\`{e}s-France\\
69676 BRON Cedex\\[3pt]
\{kaouiche, jdarmont\}@eric.univ-lyon2.fr\\[6pt]}

\resume{Avec le d\'{e}veloppement des bases de donn\'{e}es en g\'{e}n\'{e}ral et des entrep\^{o}ts
de donn\'{e}es (data warehouses) en particulier, il est devenu primordial de
r\'{e}duire la fonction d'administration de base de donn\'{e}es. L'id\'{e}e d'utiliser des
techniques de fouille de donn\'{e}es (\textit{data mining}) pour extraire des
connaissances utiles des donn\'{e}es elles-m\^{e}mes pour leur administration  est
avanc\'{e}e depuis quelques ann\'{e}es. Pourtant, peu de travaux de recherche ont \'{e}t\'{e}
entrepris. L'objectif de cette \'{e}tude est de rechercher une fa\c{c}on d'extraire des
donn\'{e}es stock\'{e}es des connaissances utilisables pour appliquer de mani\`{e}re
automatique des techniques d'optimisation des performances, et plus
particuli\`{e}rement d'indexation. Nous avons r\'{e}alis\'{e} un outil qui effectue une
recherche de motifs fr\'{e}quents sur une charge donn\'{e}e afin de calculer une
configuration d'index permettant d'optimiser le temps d'acc\`{e}s aux donn\'{e}es. Les
exp\'{e}rimentations que nous avons men\'{e}es ont montr\'{e} que les configurations
d'index g\'{e}n\'{e}r\'{e}es par notre outil permettent des gains de performance de l'ordre
de 15\% \`{a} 25\% sur une base et un entrep\^{o}t de donn\'{e}es tests.}

\abstract{With the wide development of databases in general and data warehouses
in particular, it is important to reduce the tasks that a database
administrator must perform manually. The idea of using data mining techniques
to extract useful knowledge for administration from the data themselves has
existed for some years. However, little research has been achieved. The aim of
this study is to search for a way of extracting useful knowledge from stored
data to automatically apply performance optimization techniques, and more
particularly indexing techniques. We have designed a tool that extracts
frequent itemsets from a given workload to compute an index configuration that
helps optimizing data access time. The experiments we performed showed that the
index configurations generated by our tool allowed performance gains of 15\% to
25\% on a test database and a test data warehouse.}

\motscles{Bases et entrep\^{o}ts de donn\'{e}es, Auto-indexation, Data mining, Motifs
fr\'{e}quents.}

\keywords{Databases and data warehouses, Auto-indexing, Data mining, Frequent
itemsets.}

\maketitlepage

\section{Introduction}

L'utilisation courante de bases de donn\'{e}es requiert un administrateur qui a
pour r\^{o}le principal la gestion des donn\'{e}es au niveau logique (d\'{e}finition de
sch\'{e}ma) et physique (fichiers et disques de stockage), ainsi que l'optimisation
des performances de l'acc\`{e}s aux donn\'{e}es. Avec le d\'{e}ploiement \`{a} grande \'{e}chelle
des syst\`{e}mes de gestion de bases de donn\'{e}es (SGBD), minimiser la fonction
d'administration est devenu indispensable~\cite{WEI02}.

L'une des t\^{a}ches importantes d'un administrateur est la s\'{e}lection d'une
structure physique appropri\'{e}e pouvant am\'{e}liorer les performances du syst\`{e}me en
minimisant les temps d'acc\`{e}s aux donn\'{e}es~\cite{FinkelsteinST88}. Les index sont
des structures physiques permettant un acc\`{e}s direct aux donn\'{e}es. Le travail
d'optimisation des performances de l'administrateur se porte en grande partie
sur la s\'{e}lection d'index et de vues mat\'{e}rialis\'{e}es~\cite{Agrawal2001}. Ces
structures jouent un r\^{o}le particuli\`{e}rement important dans les bases de donn\'{e}es
d\'{e}cisionnelles (BDD) tels que les entrep\^{o}ts de donn\'{e}es, qui pr\'{e}sentent une
volum\'{e}trie tr\`{e}s importante et sont interrog\'{e}s par des requ\^{e}tes complexes.

Depuis quelques ann\'{e}es, l'id\'{e}e est avanc\'{e}e d'utiliser les techniques de fouille
de donn\'{e}es (\textit{data mining}) pour extraire des connaissances utiles des
donn\'{e}es elles-m\^{e}me pour leur administration~\cite{Chaudhuri98}. Cependant, peu
de travaux de recherches ont \'{e}t\'{e} entrepris dans ce domaine jusqu'ici. C'est
pourquoi nous avons con\c{c}u et r\'{e}alis\'{e} un outil qui utilise la fouille de donn\'{e}es
pour proposer une s\'{e}lection (configuration) d'index pertinente.

Partant de l'hypoth\`{e}se que l'utilit\'{e} d'un index est fortement corr\'{e}l\'{e}e \`{a} la
fr\'{e}quence de l'utilisation des attributs correspondants dans l'ensemble des
requ\^{e}tes d'une charge donn\'{e}e, la recherche de motifs
fr\'{e}quents~\cite{agrawal93mining} nous a sembl\'{e} appropri\'{e}e pour mettre en
\'{e}vidence cette corr\'{e}lation et faciliter le choix des index \`{a} cr\'{e}er. L'outil que
nous pr\'{e}sentons dans cet article exploite le journal des transactions (ensemble
de requ\^{e}tes r\'{e}solues par le SGBD) pour proposer une configuration d'index.


\section{Conclusions et perspectives}\label{Conclusions and perspectives}

Nous avons pr\'{e}sent\'{e} dans cet article une nouvelle approche pour la s\'{e}lection
automatique d'index dans un SGBD. L'originalit\'{e} de notre travail repose sur
l'extraction de motifs fr\'{e}quents pour d\'{e}terminer une configuration d'index.
Nous nous basons en effet sur l'intuition que l'importance d'un attribut \`{a}
indexer est fortement corr\'{e}l\'{e}e avec sa fr\'{e}quence d'apparition dans les requ\^{e}tes
pr\'{e}sentes dans une charge. Par ailleurs, l'utilisation d'un algorithme
d'extraction de fr\'{e}quents tel que Close~\cite{close99is} nous permet de g\'{e}n\'{e}rer
des index mono-attribut et multi-attributs \`{a} la vol\'{e}e, sans avoir \`{a} mettre en
\oe uvre un processus it\'{e}ratif permettant de cr\'{e}er successivement des index
multi-attributs de plus en plus gros \`{a} partir d'un ensemble d'index
mono-attribut.

Nos premiers r\'{e}sultats exp\'{e}rimentaux montrent que notre technique permet
effectivement d'am\'{e}liorer le temps de r\'{e}ponse de 20\% \`{a} 25\% pour une charge
d\'{e}cisionnelle appliqu\'{e}e \`{a} une base de donn\'{e}es relationnelle (banc d'essais
TPC-R)~\cite{tpcr99}. Nous avons par ailleurs propos\'{e} deux strat\'{e}gies effectuer
une s\'{e}lection parmi les index candidats : la premi\`{e}re cr\'{e}e syst\'{e}matiquement
tous les index candidats et la deuxi\`{e}me ne cr\'{e}e que les index associ\'{e}s \`{a} des
tables dites volumineuses. La deuxi\`{e}me strat\'{e}gie apporte une meilleure
am\'{e}lioration car elle propose un compromis entre l'espace occup\'{e} par les index
(le nombre d'index cr\'{e}\'{e}s est limit\'{e} \`{a} ceux qui sont d\'{e}finis sur des attributs
de tables volumineuses) et l'int\'{e}r\^{e}t de la cr\'{e}ation d'un index (il est peu
int\'{e}ressant de cr\'{e}er un index sur une petite table).

Nous avons \'{e}galement r\'{e}alis\'{e} des tests sur un petit magasin de donn\'{e}es
d'accidentologie~\cite{accidento} auquel nous avons appliqu\'{e} une charge
d\'{e}cisionnelle ad hoc. Le gain en temps de r\'{e}ponse, de l'ordre de 14\%, est
moins important que dans le cas de TPC-R. Cela peut \^{e}tre expliqu\'{e} par le fait
que les index cr\'{e}\'{e}s par d\'{e}faut par SQL Server sont des variantes des B-arbres
et non des index \textit{bitmap} et des index de jointure en \'{e}toile, qui
seraient plus adapt\'{e}s pour un entrep\^{o}t de
donn\'{e}es~\cite{ONeil95,oneil97improved}.

Notre travail d\'{e}montre que l'id\'{e}e d'utiliser des techniques de fouille de
donn\'{e}es pour l'auto-administration des SGBD est prometteuse. Il n'est cependant
qu'une premi\`{e}re approche et ouvre de nombreuses perspectives de recherche. Une
premi\`{e}re voie consisterait \`{a} am\'{e}liorer la s\'{e}lection des index en concevant des
strat\'{e}gies plus \'{e}labor\'{e}es que l'utilisation exhaustive d'une configuration ou
l'exploitation de renseignements relativement basiques concernant la taille des
tables. Un mod\`{e}le de co\^{u}t plus fin au regard des caract\'{e}ristiques des tables
(autres que la taille), ou encore une strat\'{e}gie de pond\'{e}ration des requ\^{e}tes de
la charge (par type de requ\^{e}te : s\'{e}lection ou mise \`{a} jour), pourraient nous
aider dans cette optique. L'utilisation d'autres m\'{e}thodes de fouille de donn\'{e}es
non-supervis\'{e}es telles que le regroupement (\textit{clustering}) pourraient
\'{e}galement fournir des ensembles de fr\'{e}quents moins volumineux.

Par ailleurs, il para\^{\i}t indispensable de continuer \`{a} tester notre m\'{e}thode pour
mieux \'{e}valuer la surcharge qu'elle engendre pour le syst\`{e}me, que ce soit en
terme de g\'{e}n\'{e}ration des index ou de leur maintenance. Il est notamment
n\'{e}cessaire de l'appliquer pour des entrep\^{o}ts de donn\'{e}es de grande taille et en
tirant partie d'index adapt\'{e}s. Il serait \'{e}galement tr\`{e}s int\'{e}ressant de la
comparer de mani\`{e}re plus syst\'{e}matique avec l'outil IST d\'{e}velopp\'{e} par
Microsoft~\cite{chaudhuri97efficient}, que ce soit par des calculs de
complexit\'{e} des heuristiques de g\'{e}n\'{e}ration de configurations d'index (surcharge)
ou des exp\'{e}rimentations visant \`{a} \'{e}valuer la qualit\'{e} de ces configurations (gain
en temps de r\'{e}ponse et surcharge due \`{a} la maintenance des index).

Finalement, \'{e}tendre ou coupler notre approche \`{a} d'autres techniques
d'optimisation des performances (vues mat\'{e}rialis\'{e}es, gestion de cache,
regroupement physique, etc.) constitue \'{e}galement une voie de recherche
prometteuse. En effet, dans le contexte des entrep\^{o}ts de donn\'{e}es, c'est
principalement en conjonction avec d'autres structures physiques
(principalement les vues mat\'{e}rialis\'{e}es) que l'indexation permet d'obtenir des
gains de performance significatifs~\cite{Gupta99,AgrawalCN00,Agrawal2001}.

\bibliography{jds}
\bibliographystyle{alpha}

\end{document}